\documentclass[prl,twocolumn,tightenlines,floatfix,nofootinbib]{revtex4}
\usepackage{graphicx,amsfonts,amssymb,amsmath}

\def\ct{\tilde{c}}
\def\x{{\bf x}}
\def\GN{G_N^{(d+1)}}

\begin{document}

\title{
Black holes and universality classes of critical points
}
\author{Pavel Kovtun and Adam Ritz}
\affiliation{Department of Physics and Astronomy, University of Victoria,
             Victoria, BC, V8P 5C2 Canada}
\email{pkovtun@uvic.ca,  aritz@uvic.ca}
\date{January 2008}
\begin{abstract}
\noindent
We argue that there exists an infinite class of conformal field theories
in diverse dimensions, having a universal ratio of the
central charge $c$ to the normalized entropy density $\ct$.
The universality class includes all conformal theories which possess a
classical gravity dual according to the AdS/CFT correspondence.
From the practical point of view, the universality of $c/\ct$ provides 
an explicit test which can be applied to determine whether a given
critical point
may admit a dual description in terms of classical gravity.
\end{abstract}
\maketitle

{\it Introduction}:---Many useful quantum field theories are either
conformal field theories (CFT), or relevant deformations of a CFT.
Among other things, such field theories describe
the interactions of all known elementary particles,
the scaling behaviour near critical points in statistical mechanics,
and the degrees of freedom on the string worldsheet. 

There are several procedures via which one can arrive at an interacting CFT.
One way is by a critical limit of a statistical-mechanical model
(or, more generally, by following the renormalization group flow
to the endpoint) \cite{Wilson-Kogut}.
Another possible way to arrive at a CFT is to start with
a Lagrangian formulation of a classically conformal theory, and use extra
symmetry (such as supersymmetry) to argue that the quantum theory
must be conformally invariant as well \cite{ls}.
A prime example of such a CFT is the ${\cal N}=4$ supersymmetric Yang-Mills
theory in 3+1 dimensions.
Yet another way to arrive at a CFT is through a theory of
(quantum) gravity in Anti-de Sitter (AdS) space, and the
AdS/CFT correspondence \cite{Maldacena, GKP, Witten, Klebanov-TASI}.

In the recent years, there has been a renewed interest in CFTs
due to their emergence in quantum critical phenomena;
in particular, relativistic CFTs were proposed as the relevant language to
describe critical quantum magnets \cite{Subir-review}.
In turn, understanding CFTs in the language of classical gravity
has been useful in the studies of quantum critical transport \cite{HKSS}.
This motivates further understanding of the relation of the AdS/CFT correspondence
to quantum criticality.

In the context of the AdS/CFT correspondence, a CFT in $d$ dimensions
has a dual description in terms of string/M theory on $AdS_{d+1}\times X$,
for some compact space $X$.
The cosmological constant of $AdS_{d+1}$ takes discrete values,
determined by the quantized fluxes of various fields on $X$.
It was realized some time ago \cite{BP} that the number of
such $AdS_{d+1}\times X$ solutions in string theory is enormous:
for $d=3$ alone,
different choices of $X$ plus various fluxes give rise to an estimated
$10^{500}$ solutions -- the so-called string landscape \cite{landscape}.
It is believed that every such compactification gives rise to a CFT;
in other words, string theory allows
one to describe roughly $10^{500}$ different CFTs, or 
$10^{500}$ different universality classes in three dimensions.
It is not unreasonable to ask: does such a multitude of CFTs
include any of the real-world critical points?
For example, is there a string dual description in $AdS_4$ 
of the simplest liquid-gas critical point (the Ising CFT in $d=3$)?
A simpler question is
whether there are interesting fixed points in statistical-mechanical models
whose description (in a suitable large-$N$ limit)
may be captured by Einstein gravity in $AdS$. 
It is the purpose of this Letter to propose a 
criterion 
for whether a given CFT may have a 
dual gravitational description within 
the AdS/CFT correspondence.

{\it Central charge and entropy}:---The
central charge $c$ is a fundamental quantity which characterizes 
a given CFT. Given the standard definition of the energy-momentum tensor $T_{\mu\nu}$, we 
define $c$ via Cardy's \cite{Cardy} parametrization of
the two-point function in $d$ spacetime dimensions,
\begin{eqnarray}
  G_{\mu\nu,\alpha\beta}(x) &\equiv&  \langle T_{\mu\nu}(x) T_{\alpha\beta}(0)\rangle \nonumber\\
   &=&\frac{1}{S_d^2}\frac{c}{(x^2)^d} \left[
     \left(\delta_{\mu\alpha}-2\frac{x_\mu x_\alpha}{x^2}\right)
     \left(\delta_{\nu\beta}-2\frac{x_\nu x_\beta}{x^2}\right)\right. \nonumber\\
     && + \left.
     \left(\delta_{\mu\beta}-2\frac{x_\mu x_\beta}{x^2}\right)
     \left(\delta_{\nu\alpha}-2\frac{x_\nu x_\alpha}{x^2}\right) \right.\nonumber\\
    && -\left. \frac{2}{d}\delta_{\mu\nu}\delta_{\alpha\beta}
     \right]\!\!,
\label{eq:c-defined}
\end{eqnarray}
with $S_d=2\pi^{d/2}/\Gamma(d/2)$ the volume of the unit  $(d-1)$-sphere.

The value of $c$ is one way to measure
the number of degrees of freedom in a CFT.
A more conventional way to measure the number of degrees of freedom
in any system is to heat it up and study its entropy as a function
of temperature.%
\footnote{
   The formulas are written for $d$ Euclidean dimensions, but
   the discussion can be repeated almost verbatim
   for Minkowski CFTs with $d{-}1$ dimensional space plus time.
   For a Euclidean theory, heating up means compactifying
   one of the $d$ dimensions with periodic boundary conditions for
   bosons and anti-periodic for fermions.
   The temperature $T$ is then the inverse compactification radius.
}
In a $d$ dimensional CFT, the entropy density
$s{=}S/V$ (we take $V\to\infty$) is proportional to $T^{d-1}$
because $T$ is the only dimensionful parameter.
The dimensionless proportionality
coefficient measures the number of degrees of freedom in the system.
We write this relation as
\begin{equation}
  s = \ct\, \frac{\Gamma(d/2)^3}{4\pi^{d/2}\Gamma(d)}
  \left(\frac{4\pi}{d}\right)^d \frac{d{-}1}{d{+}1}\; T^{d-1},
\label{eq:ct-defined}
\end{equation}
which defines the ``normalized entropy density'' $\ct$.
The dimension-dependent normalization factor is simply a convenient convention.
In $d=2$ dimensions, $c$ and $\ct$ are related by conformal symmetry
\cite{BCN, Affleck}, so that
(given our conventions)
\begin{equation}
   d=2\;\;\;\; \Longrightarrow \;\;\;\; c = \ct\,, 
  \label{2D}
\end{equation}
or $s=\pi c T/3$.
It is essentially this relation between $c$ and $\ct$
in two-dimensional CFTs
which allowed for the string theoretic calculation
of black hole entropy \cite{Strominger-Vafa}.
In three dimensions, the ratio $c/\ct$ has been computed
in the critical $O(N)$ sigma model at large $N$ \cite{Subir},
with the result:
\begin{equation}
  \frac{c}{\ct} = \frac{5\pi^4}{2^4 3^3\,\zeta(3)} = 0.937909...  \label{3D}
\end{equation}
However, a general relation between $c$ and $\ct$ in dimension
three and higher
is unknown, and remains an important open problem.
We will now show that in all $d\geqslant3$ dimensional CFTs
which admit a dual gravitational description
via the AdS/CFT correspondence, the central charge is
equal to the normalized entropy density,
\begin{equation}
   \mbox{AdS/CFT}\;\;\; \Longrightarrow\;\;\;\; c=\ct\,.
\label{eq:main}
\end{equation} 
This is the main result of the paper to be derived below,
and will be discussed further at the end of this Letter.

{\it Entropy in AdS/CFT}:---Within the AdS/CFT correspondence, a $d$-dimensional CFT
at finite temperature is described by a $D$-dimensional black hole metric
\begin{equation}
   ds^2 = \frac{r^2}{L^2}\left(f(r)dx_0^2+d\x^2\right) +
          \frac{L^2}{r^2}\frac{dr^2}{f(r)} +
          L_X^2 ds_X^2\,.
\label{eq:metric}
\end{equation}
Here $x=(x_0,\x)$ is $d$-dimensional,
$L^2$ sets the value of the cosmological constant,
$f(r)=1-(r_0/r)^d$ where $r=r_0$ is the horizon,
and $ds_X^2$ is the metric on $X$.
The temperature of the CFT is $T=r_0 d/(4\pi L^2)$, and
the zero-temperature limit corresponds to $r_0\to0$.
The entropy is
proportional to the $D{-}2$ dimensional area of the horizon,
$S=A_{D-2}/4G_N$
where 
$G_N$ is the $D$-dimensional Newton's constant.
Dividing by the (infinite) $(d{-}1)$--volume $V$,
one finds for the entropy density,
\begin{equation}
  s = \frac{1}{4\GN} \left(\frac{4\pi L}{d}\right)^{d-1} T^{d-1},
\label{eq:entropy}
\end{equation}
where the $(d{+}1)$--dimensional Newton's constant is
$1/\GN = L_X^{D-(d+1)} Vol(X)/G_N$.

{\it Central charge in AdS/CFT}:---To find the central charge $c$ in dimension $d\geqslant3$,
one can use either the position or momentum space representation
of the correlation function (\ref{eq:c-defined}) (at $T=0$). 
A convenient momentum space representation is \cite{KS}
\begin{eqnarray}
  G_{\mu\nu,\alpha\beta}(k) &=& 
  \Big(\left. P_{\mu\alpha}P_{\nu\beta} + P_{\mu\beta}P_{\nu\alpha} \right. \nonumber\\
  & & -\left.
  \frac{2}{d-1} P_{\mu\nu} P_{\alpha\beta}
  \right)
  G(k^2),
\end{eqnarray}
where $P_{\mu\nu}=\delta_{\mu\nu} - k_\mu k_\nu/k^2$.
The central charge is related to $G(k^2)$ by
\begin{equation}
  \frac{c}{(x^2)^d} = \frac{d+1}{d-1} S_d^2
  \int \frac{d^dk}{(2\pi)^d}\, e^{ikx} G(k^2)\,.
  \end{equation}
Choosing $k$ along $x_i$ with $i\,{\neq}\,1$ or $2$, one has $G_{12,12}(k)=G(k^2)$,
and therefore it suffices to evaluate $G_{12,12}$ to find the
central charge.
According to the AdS/CFT prescription, $G_{12,12}$ is given by the
second variation of the gravitational action with respect to
the boundary value of the metric perturbation $h_{12}$.
The $h_{12}$ perturbation decouples from all other perturbations, and 
obeys the equation of motion coming from the action of a massless
scalar in the $AdS_{d+1}$ background,
\begin{eqnarray}
  && \int\! d^{d+1}\!x \sqrt{g}\,\big(R-\Lambda\big) +
   2\!\int\! d^dx \sqrt{g_B}\, K \nonumber\\
    && =-\frac12
   \int\! d^{d+1}\!x \sqrt{g}\, g^{\mu\nu}\partial_\mu\phi\, \partial_\nu\phi
   +\dots
\label{eq:action}
\end{eqnarray}
Here $\phi\equiv h_2^1$ (we require $d\geqslant3$),
the cosmological constant is $\Lambda=-d(d{-}1)/L^2$, the second
term in the action is the standard Gibbons-Hawking boundary term,
and contact boundary terms are dropped on the right-hand side
of (\ref{eq:action}). 
The two-point correlation function for the massless scalar
can be evaluated using the standard AdS/CFT prescription
either in momentum space, following \cite{GKP}, or in position
space \cite{Witten,FMMR}.
Restoring the overall factor of $1/16\pi\GN$ in front of the action
(\ref{eq:action}), one finds for the central charge,
\begin{equation}
  c = \frac{d+1}{d-1} \frac{L^{d-1}}{4\pi\GN} 
      \frac{\Gamma(d+1)\pi^{d/2}}{\Gamma(d/2)^3}\,,
\end{equation}
in agreement with \cite{lt}.
Comparing this to $\ct$ as found from (\ref{eq:entropy}),
we arrive at our main result in Eq.~(\ref{eq:main}),
$c/\ct=1$. For ${\cal N}=1$ supersymmetric CFTs in $d=4$, this relation was 
discussed earlier in Ref.~\cite{nt}.

{\it Discussion}:---What we have shown is that every CFT in dimension $d\geqslant3$
which has an $AdS_{d+1}$ gravity dual description must have a central charge
equal to the normalized entropy density.
More precisely, this equality should hold up to corrections which
vanish in the limit in which a classical gravitational description in $AdS$ is valid, e.g.\ at 
large $N$ and large 't Hooft coupling for ${\cal N}=4$ SYM and variants thereof.
Furthermore, we should note that the 
reduction in (\ref{eq:action}) clearly fails for $d\leqslant 2$ as Einstein gravity then has no propagating degrees of freedom; nonetheless 
conformal invariance is sufficient in $d=2$ to ensure $c{=}\ct$
for all CFTs regardless of whether they admit a dual gravitational description
in $AdS_3$. It is natural to ask if there are ``conventional'' CFTs which
also have $c=\ct$. It is obvious that $c$ cannot be equal to $\ct$ in non-interacting
(or weakly interacting) theories when $d$ is odd.
This is because in a free theory $\ct$ is proportional to $\zeta(d)$,
while $c$ contains no such irrational factors. In even dimensions, this is no longer the case and
one may wonder whether free theories may exist which satisfy $c=\ct$. 
We can test this in four dimensions where, for a free theory with $n_v$ vector, $n_f$ fermionic and
$n_s$ scalar degrees of freedom, we find that 
\begin{equation}
  \left(\frac{c}{\tilde{c}}\right)^{\rm free}_{4d} =
  \frac{3}{8} \;
  \frac{n_s+ \frac{3}{2} n_f+ 12n_v}{n_s + \frac{7}{8}n_f + 2n_v}, \label{free}
\end{equation}
so that $3/8 \leqslant c/\tilde{c}\leqslant 9/4$.
Note that $c/\tilde{c} = 3/4$ for the free limit of ${\cal N}=4$ SYM
consistent with the known difference of
the entropy (and thus $\ct$) at strong and weak coupling \cite{Klebanov-TASI}.
Importantly, we observe from this
relation that free theories do exist which satisfy $c=\tilde{c}$, 
provided they contain vector degrees of freedom, such that $2n_s+n_f=8n_v$; the
free limit of QED with two flavors is a simple example. 
Thus we conclude that, at least
in $d=4$, the relation $c=\ct$ is necessary but not sufficient for a given
CFT to possess a gravity dual.
However, when $d$ is odd, it is tempting to conjecture that the condition
$c=\ct$ is not only necessary, but also sufficient for a given CFT
to have a gravity dual.

Let us also point out that 
a criterion similar to (\ref{eq:main}) is known for $d=4$ CFTs.
In four dimensions, there are two central charges,
commonly denoted by $c$ and $a$ 
(which characterize the response to two different curvature
invariants when the CFT is placed in curved space).
It turns out that the AdS/CFT formulation implies 
(assuming a certain choice of normalization) that
$c=a$ in the limit that the classical gravitational description is valid
\cite{Henningson-Skenderis,Gubser}.
The condition $c=a$ has been considered as a means of classification \cite{anselmi}
and is necessary for a $d=4$ CFT to have a
dual gravity description in the appropriate large-$N$ limit. However,  it also is not sufficient because
there are examples (such as ${\cal N}=4$ super Yang-Mills)
where $c=a$ holds at both strong and weak coupling \cite{Henningson-Skenderis,Gubser},
while the gravitational description is only valid at strong coupling.
The condition $c=\ct$ is clearly stronger than $c=a$ because
{\it i)} it applies in any dimension $d\geqslant3$, not just in $d=4$, and
perhaps more importantly
{\it ii)} it is in principle capable of making a distinction between
strongly and weakly coupled theories because $\ct$ is not
protected by supersymmetry.

In physical terms, the condition $c/\ct=1$ is a real-space counterpart of the
relation $\eta/s=1/4\pi$, where $\eta$ is the shear viscosity
of any field theory with a dual $AdS$ gravity description \cite{KSS, Buchel}.
Indeed, the Kubo formula for $\eta$ relates the shear viscosity
to the thermal real-time correlation function $G_{12,12}(k)$
at small timelike momentum.
On the other hand, the central charge is related to $G_{12,12}(k)$
at large spacelike momentum, where the effects of temperature
do not matter. This analogy can be made more transparent if we
trade $c$ for the absorption cross-section $\sigma(\omega)$ 
(in $D$-dimensional Planck units) for graviton scattering
by the appropriate gravitational background \cite{Klebanov-TASI}
and contrast $\eta/s$ with the high-frequency limit of
$\sigma(\omega)/s\propto c/\ct$ at temperature $T$.\,\footnote{
  The fact that
  $c/\ct=1$ implies that 
  $\sigma/(s G_N) = f(d) (\omega/T)^{d-1}$, where the constant $f(d)$ 
  is purely a function of the spacetime dimension of the CFT.
  For $d=3,4,6$, the constant $f(d)$ can be found from the
  $AdS$ duals of M2, D3, and M5 branes, using the results
  \cite{Klebanov-TASI} for $\sigma(\omega)$.
} 
Indeed, the dual gravitational perturbation $\phi=h_2^1$ behaves as
a massless scalar in Eq.~(\ref{eq:action}), regardless of the
temperature \cite{KSS}. 
However, within AdS/CFT, the ratio $\eta/s$ apparently defines a wider class
than $c/\ct$ because $\eta/s=1/4\pi$ for both CFTs
{\it and} relevant deformations of CFTs. 
In addition, $\eta/s=1/4\pi$ applies universally for any $d\geqslant3$,
without dimension-dependent normalization factors.
However, from the point of view of finding a possible gravity
dual for a given CFT, the condition $c/\ct=1$ has a significant
advantage over $\eta/s=1/4\pi$ because it involves only
equilibrium quantities which are easier to compute than 
real-time response functions. 

Focusing on 3-dimensional systems, relevant to real-world critical points, we find it interesting that the 
large-$N$ result (\ref{3D}) for $c/\ct$ in the $O(N)$ model is numerically very close to one, the value required 
for the existence of a gravity dual. With regard to the proposal of Klebanov and Polyakov \cite{kp} --
that the large-$N$ dual 
is a higher-spin gauge theory in AdS -- this result amounts to a prediction for the bulk spin-two sector and implies a 
(small) quantitative difference with pure Einstein gravity. 
At the opposite end of the spectrum, the 
critical Ising model corresponds to $N=1$ and it would clearly be interesting to see 
if $1/N$ corrections \cite{petkou},
known to be generically rather large \cite{Zinn-Justin}, were to modify Eq.~(\ref{3D}) bringing $c/\ct$ closer to
one ($c/\ct$ could also be computed directly at the Wilson-Fisher fixed point
using the epsilon expansion). 
However, the possibility of $c/\ct=1$ in this case would go against the general
expectation that the critical
Ising model has too few degrees of freedom to possess a classical gravity dual. 
This is because every CFT with a classical gravity dual has a finite-volume phase
transition \cite{Witten-BH} as a function of $T$, and therefore must
have an infinite number of degrees of freedom.
Nevertheless, one hopes that suitable limits exist in which these CFTs
are close to real-world examples. It would be interesting to investigate more generally 
the corrections to $c/\ct = 1$ arising from quantum corrections to classical gravity.

Finally, we have seen that $AdS$ gravity provides us
with a multitude of non-Gaussian fixed points with exactly the same value of $c/\ct$.
However, it is far from obvious how these fixed points are related to each other;
in particular, they have different symmetries. While Monte Carlo simulations should be
tractable in many cases, we are not aware of numerical results for
$c/\ct$ at non-Gaussian fixed points in any three-dimensional lattice model.
Thus, going beyond the classical gravity approximation, 
as a related question we may ask: are there non-Gaussian fixed points
in three dimensions that share the same value of $c/\ct$ (not necessarily equal to one)?
A positive answer would suggest a novel notion of universality,
which is not related to symmetry, but may be related to (quantum) gravity in Anti-de Sitter  space.

\acknowledgments
We are grateful to Subir Sachdev for helpful discussions, and to Sean Hartnoll, Chris Herzog, Anastasios Petkou, Kostas Skenderis, Tadashi Takayanagi, and Edward Witten 
for comments on the manuscript. This work was supported in part by NSERC of Canada.

\end{document}